\documentclass[twocolumn,twoside,slac_two]{revtex4}
\usepackage{graphicx}
\usepackage{fancyhdr}
\begin{document}

\title{Neutrino Masses and Mixing -- Theory}

%

\author{Andr\'e de Gouv\^ea}
\affiliation{Northwestern University, Illinois, United States of America}

\begin{abstract}
In this talk I review what we know and don't know about neutrinos, neutrino masses and lepton mixing. I also discuss the importance of the discovery that neutrinos have nonzero masses, and illustrate how little is currently known about the physics behind them. 
\end{abstract}

\maketitle

\thispagestyle{fancy}


\section{What We Learned About Neutrinos}

Over the past decade, our understanding of neutrinos changed dramatically. After decades of confusion, it is now established that neutrinos change flavor after propagating a finite distance. The ``rate of change'' depends on the neutrino energy, the distance between the neutrino source and the neutrino detector, and, in some cases, the medium through which the neutrinos propagate. For a detailed summary of all data, see, for example, \cite{deGouvea:2004gd,GonzalezGarcia:2007ib}. For the most recent results and updates see \cite{neutrino2008,Franco}.

The only consistent explanation to all long-baseline neutrino oscillation data is to postulate that, just like all other fermions in the standard model, neutrinos have non-zero, distinct masses and that, like the quarks, leptons mix. This means that the charged-current weak interactions are not ``aligned'' with the charged and neutral lepton mass eigenstates. In this case, the observed neutrino flavor-change is a consequence of neutrino oscillations and the data constrain the neutrino mass-squared differences and different combinations of the elements of the lepton mixing matrix, often referred to as the Pontecorvo-Maki-Nakagawa-Sakata (PMNS) matrix.

It is often the case that lepton mixing is described in the weak basis where the charged-lepton mass matrix is diagonal (with eigenstates $e,\mu,\tau$) so that the PMNS matrix $U$ relates the neutrino weak-eigenstates $\nu_\alpha$, $\alpha=e,\mu,\tau$ and the neutrino mass eigenstates $\nu_i$ (with mass $m_i$), $i=1,2,3$ (for now and until further notice, I'll assume there are three neutrino species):
\begin{equation}
\nu_{\alpha}=U_{\alpha i}\nu_i.
\end{equation}
The elements of $U$ are often parameterized as prescribed by the Particle Data Group \cite{Amsler:2008zzb} by three mixing angles $\theta_{12},\theta_{13},\theta_{23}$ and three CP-odd phases $\delta,\alpha_1,\alpha_2$. If the neutrinos are Dirac fermions, the convention is such that only the phase $\delta$ is a physical observable.

\begin{figure}[ht]
\centering
\includegraphics[width=80mm]{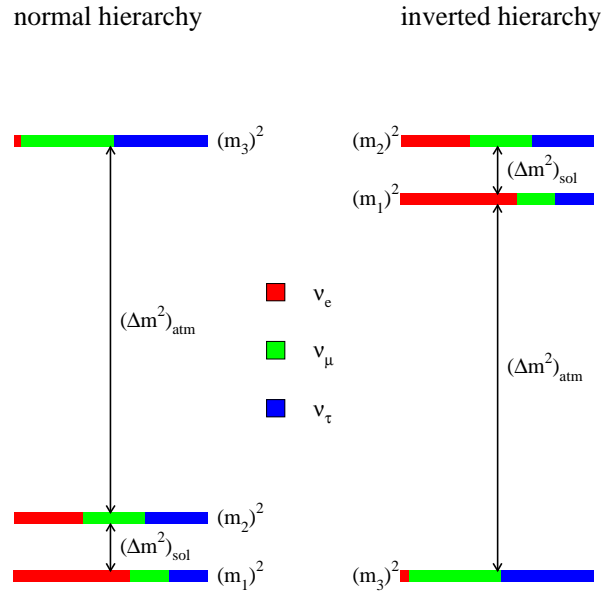}
\caption{Cartoon of the relationship between neutrino mass and flavor eigenstates, for both the normal and inverted neutrino mass hierarchy. The fractions of the different horizontal bars with a specific color are representative of the different $|U_{\alpha i}|^2$. From \cite{deGouvea:2004gd}, where one can go for more details.} \label{3nus}
\end{figure}

In order to match mixing parameters to data, it is imperative to properly define the neutrino mass-eigenstates. This is done in the following standard if somewhat unusual way. Given three neutrinos, one can define three mass-squared differences, two of which are independent. States $\nu_1$ and $\nu_2$ define the smallest (in magnitude) mass-squared difference and we further impose $m_1^2<m_2^2$. State $\nu_3$ is the left-over state which may be either heavier or lighter than $\nu_1$ and $\nu_2$. Note that while $\Delta m^2_{12}\equiv m_2^2-m_1^2$ is positive-definite, $\Delta m^2_{13}$ (and $\Delta m^2_{23}$) can have either sign. A so-called normal (inverted) neutrino mass hierarchy is associated to $\Delta m^2_{13}>0$ ($\Delta m^2_{13}<0$). Both mass hierarchies are depicted in Fig.~\ref{3nus}. For a careful discussion of the neutrino oscillation parameters, see  \cite{deGouvea:2008nm}.

Current data constrain, according to \cite{Schwetz:2008er},

\begin{eqnarray}
\Delta m^2_{12}= (7.65^{+0.23}_{-0.20}) \times10^{-5}~{\rm eV^2}, &  \sin^2\theta_{12}=0.30^{+0.02}_{-0.02}, \nonumber \\
|\Delta m^2_{13}|=(2.40^{+0.12}_{-0.11})\times 10^{-3}~{\rm eV^2}, & \sin^2\theta_{23}= 0.50^{+0.07}_{-0.06},\nonumber \\
\sin^2\theta_{13}<0.040~~(2\sigma~{\rm bound}), & \delta\in [0,2\pi],  \label{measurements}
\end{eqnarray}
 while the sign of $\Delta m^2_{13}$ is unconstrained. The so-called Majorana phases $\alpha_1$ and $\alpha_2$, if physically observable, are currently completely unconstrained.
 
 Before proceeding, it is important to appreciate that neutrino data do not require ``more'' non-standard neutrino properties, including anomalous magnetic moments, non-standard neutrino interactions, or new neutrino states. Indeed, many of these are currently constrained by neutrino oscillation data, sometimes significantly \cite{GonzalezGarcia:2007ib}.
 
\section{What We Know We Don't Know}

While neutrino data have revealed new lepton properties (neutrino mass-squared differences and leptonic mixing angles), there are still several ``known unknowns'' that need to be uncovered by next-generation neutrino experiments. These are not new, exotic neutrino properties that may or may not be present. Here I'll discuss questions to which, given the fact that neutrinos have mass, there are well-defined, always relevant, answers.

\subsection{Unknown Oscillation Parameters}

A quick inspection of the results listed in Eq.~\ref{measurements} reveals that some of the oscillation parameters are either unknown or only poorly constrained. In more detail:
\begin{enumerate}
\item $\sin^2\theta_{13}$ is only constrained to be less than several percent. $\sin^2\theta_{13}$ is also the magnitude of the $U_{e3}$ matrix element, and is equal to the probability that a $\nu_3$ is measured as what we normally refer to as an electron neutrino  $\nu_e$. Hence, current data cannot tell whether $\nu_3$ is a linear combination of only $\nu_{\mu}$ and $\nu_{\tau}$ or whether it also has a $\nu_e$ component. The answer to this question may prove valuable when it comes to testing different neutrino mass models (for an overview, see \cite{theory_review}).
\item The sign of $\cos2\theta_{23}$ is virtually unconstrained. This means that we can't tell whether the $\nu_3$ state has a larger $\nu_{\tau}$ or $\nu_{\mu}$ component, or whether the two components ($U_{\tau3}$ and $U_{\mu3}$) have the same magnitude. The answer to this question, which has no parallel in the quark sector, should help guide theoretical understanding of the origin of neutrino masses and the structure of lepton mixing  \cite{theory_review}.
\item The CP-odd phase $\delta$ is completely unconstrained. $\delta\neq 0,\pi$ indicates that CP-invariance is violated in neutrino oscillations, ${\it i.e.}$, $P(\nu_{\alpha}\to\nu_{\beta})\neq P(\bar{\nu}_{\alpha}\to\bar{\nu}_{\beta})$. I'll comment more on this momentarily. The importance of understanding CP-invariance violating phenomena in fundametal physics has been emphasized over the past several decades. Here I'll only mention that, so far, all observed CP-invariance violating phenomena (in kaon and $B$-meson mixing and decay) are parametrized by one CP-odd Lagrangian parameter: the phase in the quark mixing matrix. Neutrino oscillations provide the first opportunity to observe a CP-invariance violating phenomenon that probes a guaranteed new source of CP-invariance violation.   
\item The neutrino mass hierarchy, here parameterized by the sign of $\Delta m^2_{13}$ is unconstrained. Different neutrino mass hierarchies point to potentially very different neutrino mass theories  \cite{theory_review} and potentially qualitative neutrino mass spectra. In the case of an inverted neutrino mass hierarchy, for example, two of the neutrino masses are guaranteed to be almost degenerate: $m_2-m_1\ll m_2,m_1$. This featured is not shared by any other standard model entities, except for those which are related by approximate symmetries. Examples of composite states include the pions and nucleons, which are related by isospin symmetry. At the fundamental level, on the other hand, the $W$ and $Z$-boson masses are related by custodial symmetry. If the neutrino mass spectrum is inverted, we will be very tempted to believe that there is some new approximate symmetry that relates at least some of the different lepton-doublet fields.  
\end{enumerate}
Obtaining the answers to these four questions is the current driving force behind the next-generation neutrino oscillation experimental program. Whether one can obtain a satisfactory answer to any of them depends on the magnitude of $|U_{e3}|^2$. Ultimately, the goal of next and next-next-generation experiments is {\sl not} to precisely measure all neutrino oscillation parameters, but rather to test the three-massive-neutrino paradigm. While the three-massive-neutrino paradigm works very well there is lots of room for expansion. Only after ``over-constraining'' the parameters will we be convinced that the neutrino sector has been properly understood.

Before proceeding, I'll schematically describe how CP-invariance violation manifests itself in neutrino oscillations. The amplitude related to the probability that, say, a neutrino produced as a $\nu_{\mu}$ wil be detected as a $\nu_e$ is given by (in vacuum)
\begin{equation}
A_{\mu e}=U_{e2}^*U_{\mu2}\left(e^{i \Delta_{12}}-1\right)+U_{e3}^*U_{\mu3}\left(e^{i \Delta_{13}}-1\right),
\end{equation}
where $\Delta_{1i}=\frac{\Delta m^2_{1i}L}{2E}$, $i=2,3$, $L$ is the distance travelled by the neutrino and $E$ its energy. The amplitude for the CP-conjugate process is
\begin{equation}
{\bar A}_{\mu e}=U_{e2}U_{\mu2}^*\left(e^{i\Delta_{12}}-1\right)+U_{e3}U_{\mu3}^*\left(e^{i \Delta_{13}}-1\right).
\end{equation}
Above, it was assumed that $U$ is a unitary matrix. $|A|^2\neq |\bar{A}|^2$ if  (a) there are non-trivial ``weak phases'', {\it i.e.}, the relative phase of the different $U_{ei}U_{\mu i}$ must be non-zero; (b) there are non-trivial ``strong-phases'', {\it i.e.}, $\Delta_{12}\neq\Delta_{13}\neq 0$. This is very similar to the conditions for observing CP-invariance violation in meson decays. Condition (a) translates into $\delta\neq 0,\pi$ {\sl and} $\theta_{13}\neq 0$, while condition (b) translates into $L\neq 0$ and $\Delta m^2_{12},\Delta m^2_{13}\neq0$. All (b) requirements can be easily met by appropriately choosing the baseline $L$. Our ability to probe CP-violation depends on whether $U_{e3}$ (or $\theta_{13}$) is non-zero (and large enough). This is part of the reason next-generation experiments discuss most their sensitivity to $\theta_{13}$. It determines our ability to study CP-invariance violation in neutrino oscillations and also drives the strategy behind addressing most of the issues numbered above. 

\subsection{How Light is the Lightest Neutrino?}

Neutrino oscillation experiments are only sensitive to neutrino mass-squared differences. Knowledge of these does not allow one to determine the neutrino masses themselves. This uncertainty can be parameterized in terms of the lightest neutrino mass, $m_{\rm ltest}$. Once that is known -- along with the neutrino mass hierarchy and the mass-squared differences -- all neutrino masses are determined:
\begin{eqnarray}
& m_1^2=m_{\rm ltest}^2; & m_1^2 = m_{\rm ltest}^2 - \Delta m^2_{13}, \nonumber \\
& m_2^2=m_{\rm ltest}^2+\Delta m^2_{12}; & m_2^2= m_{\rm ltest}^2 - \Delta m^2_{13} + \Delta m^2_{12}, \nonumber \\
& m_3^2=m_{\rm ltest}^2+\Delta m^2_{13}; & m_3^2=m_{\rm ltest}^2. \nonumber \\
\end{eqnarray}
The left-hand (right-hand) expressions apply in the case of a normal (inversted) hierarchy. Different values of $m_{\rm ltest}$ lead to qualitatively different neutrino spectra. If, for example, $m^2_{\rm ltest}\gg |\Delta m^2_{13}|$, all neutrino masses are quasi-degenerate: $|m_i-m_j|\ll m_i$ for all $i,j$.  On the other hand, if  $m^2_{\rm ltest}\ll \Delta m^2_{12}$ and the neutrino mass hierarchy is normal the neutrino masses are hierarchical: $m_1\ll m_2\ll m_3$. Current constraints on $m_{\rm ltest}$ and prospects for the future were discussed in detail in this conference \cite{absolute_mass}.

\subsection{Are Neutrinos Majorana Fermions?}

Unlike all other standard model fermions, neutrinos have zero electric charge. This means that, unlike all other standard model fermions, they may be Majorana fermions. Majorana fermions are their own antiparticles in the same sense that the photon or the neutral pion are their own antiparticles. Another way to contrast a massive Majorana fermion from a Dirac one (all charged leptons and quarks are Dirac Fermions) is that a Dirac fermion field describes four degrees of freedom, while a Majorana fermion field describes only two. 

It is instructive to discuss a well-known example. If the electron did not have mass, we could talk about the left-handed helicity electron state, which would be the state destroyed by the left-handed chiral electron field. The CPT-theorem dictates that, associated to this state, there is the right-handed positron state, which is created by the the same left-handed chiral electron field. If the electron has mass, however, helicity is not a conserved quantum number. Indeed, it depends on the reference frame and can be ``changed'' with a Lorentz boost. In other words, we can go to the electron rest frame and, along with one ``helicity'' (say, the spin up projection) there must also exist a different state with the opposite helicity but the same electric charge (in this case the spin down projection). This right-handed electron is destroyed by a new chiral field, the right-handed chiral electron field. Its CPT-conjugate is the left-handed positron, which is created by the same field. 

One can repeat the same story for the neutrinos. The CPT-partner of the left-handed neutrino is referred to as the left-handed antineutrino. Once neutrino masses are included, the left-handed neutrino must also have a ``Lorentz-partner.'' Unlike the electron, however, the neutrino has no electric charge, and there is hence a choice. One can either introduce a new right-chiral field (the right-handed neutrino field) that partners up with the left-chiral one to give neutrinos a mass or postulate that the right-handed antineutrino is both its CPT-partner and its Lorentz-partner. In the former case, each massive neutrino is described by four degrees of freedom: the left and right-handed neutrinos and the left and right-handed antineutrinos. In the latter case, only two: a left-handed state (usually called a neutrino) and a right-handed one (usually called an antineutrino).

Above, we flagged the neutrino as special because it had no electric charge. In order for it to be a massive Majorana fermion, it cannot possess {\sl any} quantum numbers {\sl after} electroweak symmetry is broken. In the standard model, however, the neutrinos does possess a quantum number. Neutrinos are charged under a global, non-anomalous $U(1)_{B-L}$ symmetry. Hence, if the neutrinos are massive Majorana fermions, this symmetry cannot be exact and lepton number minus baryon number violation is guaranteed to occur. Searches for lepton number violation provide the deepest probes for the Majorana nature of the neutrinos. 

It is important to appreciate that the issue in question -- whether neutrinos are Majorana fermions -- can only be addressed if the neutrinos have mass. In the case of massless neutrinos, helicity is a good quantum number and only two neutrino degrees (the left-handed neutrino and the right-handed antineutrino) of freedom interact with other standard model particles regardless of whether right-handed chiral neutrino fields exist. Furthermore, nothing ``mixes'' the left- and right-handed neutrino states and $U(1)_{B-L}$ can be a conserved global symmetry.All of these imply that any observable that is sensitive to the Majorana nature of the neutrino disappears in the limit when neutrino masses are zero. The amplitudes for such processes are proportional to $m_{\nu}/E$, where $m_{\nu}$ is some combination of  neutrino masses and $E$ is the neutrino energy associated to the process in question. Since neutrino masses are tiny, all observables capable of addressing the Majorana versus Dirac question are very, very rare. 

\begin{figure}[ht]
\centering
\includegraphics[width=80mm]{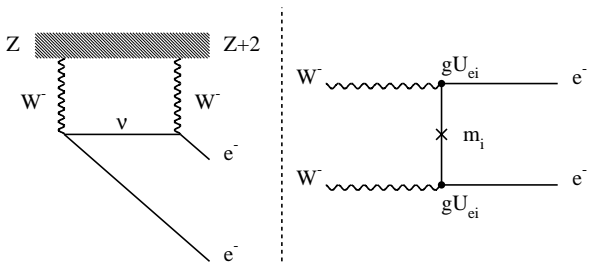}
\caption{LEFT -- Diagram contributing to neutrinoless double-beta decay. The hatched region indicates the ``nuclear physics'' part of the process. RIGHT -- The high-energy ``core'' of neutrinoless double-beta decay, $W^-+W^-\to e^-+e^-$ via $\nu_i$ exchange ($i=1,2,3$, $m_{\nu_i}\equiv m_i$). Here, $g$ is the weak coupling, $U$ is the lepton mixing matrix, and the cross indicates a fermion mass insertion. From \cite{deGouvea:2004gd}. \label{fig:0nubb}}
\end{figure}

The best probe of the Majorana nature of the neutrino and of the conservation of lepton number is neutrinoless double-beta decay. This was discussed in some detail in this conference \cite{absolute_mass}. It consists of the decay
\begin{equation}
Z\to(Z+2)e^-e^-,
\label{0nubb}
\end{equation}
 where $Z, Z+2$ stand for  nuclei with atomic numbers $Z, Z+2$ respectively. It is easy to see that Eq.~\ref{0nubb} violates lepton number by two units. Very naively, this process occurs when two virtual $W$-bosons are emitted ``at the same time,'' as depicted in Fig.~\ref{fig:0nubb}. The $W$-bosons manifest themselves as an electron-antineutrino pair. Neutrinoless double-beta decay occurs when the two antineutrinos, instead of manifesting themselves as real states, ``annihilate.'' This can only occur if neutrinos are their own antiparticles.

The amplitude for the process above is proportional to the neutrino propagator. Due to the nature of the $W$-boson interaction, the contributing part of the neutrino propagator (for  a recent pedagogical discussion on Feynman diagram computations with Weyl and Majorana fermions see \cite{Dreiner:2008tw}) is proportional to the neutrino mass and
\begin{equation}
A_{0\nu\beta\beta}\propto g^2\sum_i U_{ei}U_{ei} \frac{m_i}{Q^2}\equiv \frac{g^2}{Q^2}m_{ee}.
\end{equation}
where $Q$ is representative of typical neutrino energies and we assumed that these are much larger than the neutrino masses. $m_{ee}$ is an effective neutrino mass to which neutrinoless double-beta decay experiments are sensitive, and agrees with the $ee$-elelment of the Majorana neutrino mass matrix in the weak-basis where the charged-current and the charged-lepton mass matrices are diagonal. $m_{ee}$ was discussed in detail in \cite{absolute_mass}. Here it suffices to say that it is a weighted average of the different neutrino masses, and that the weights are complex. Not only can $m_{ee}$ vanish, but it also depends, in principle, on the Majorana phases $\alpha_1$ and $\alpha_2$.

\section{Neutrinos Have Mass - So What?}

In the previous sections I argued that nonzero neutrino masses have been discovered, along with lepton mixing. It remains to introduce these new ingredients into the standard model of strong and electroweak interactions. Before discussing this, it is worthwhile to inspect what we have uncovered regarding neutrino masses, and how they compare with the rest of the standard model. 

\begin{figure}[ht]
\centering
\includegraphics[width=80mm]{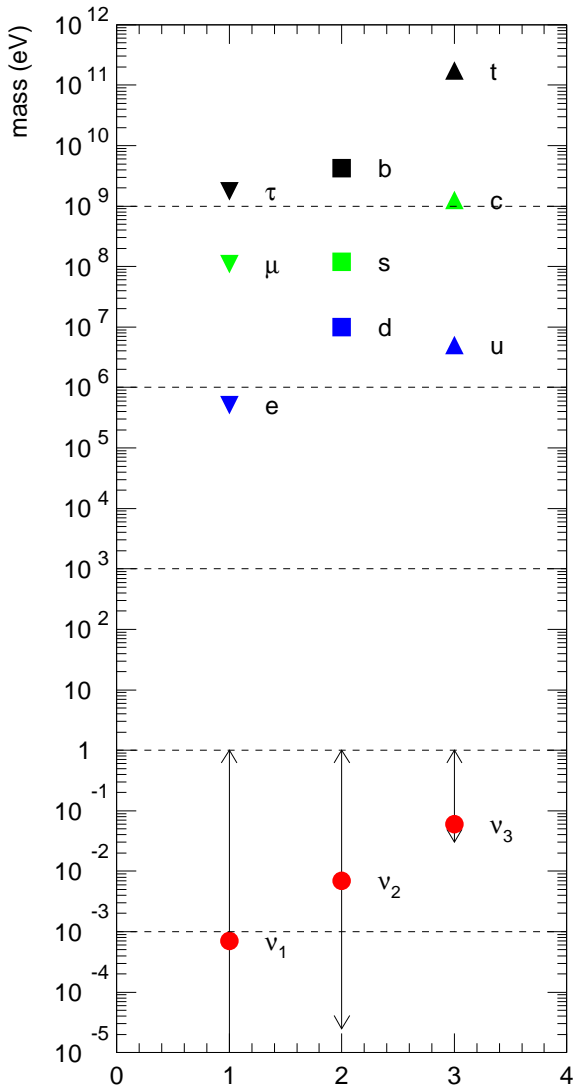}
\caption{Standard model fermion masses. For the neutrino masses, the normal mass hierarchy was assumed, and a loose upper bound $m_i<1$~eV, for all $i=1,2,3$ was imposed. From \cite{deGouvea:2004gd}. \label{fig:masses}}
\end{figure}

Neutrino masses, while non-zero, are really tiny. Fig.~\ref{fig:masses} depicts the masses of all standard model fermions, including the neutrinos. Two features stand out immediately. First, neutrino masses are at least six orders of magnitude less than the electron mass. Readers are reminded that the electron mass itself is already over 100 times smaller than the muon mass and tiny compared to the weak scale, around 100~GeV. Second, to the best of our knowledge, there is a ``gap'' between the largest allowed neutrino mass and the electron mass, in contrast with the fact that, in the charged-fermion part of the mass-space one  encounters a new mass every order of magnitude or so. We don't know why neutrino masses are so small or why there is such a large gap between the neutrino and the charged fermion masses. We suspect, however, that this may be Nature's way of telling us that neutrino masses are ``different.'' This suspicion is only magnified by the fact that neutrinos may be Majorana fermions. 

On top of these  theoretical hunches there is the fact that, in the standard model, neutrinos were predicted to be exactly massless. The discovery of neutrino masses, hence, qualifies as the first concrete instance where the standard model failed. More important is the fact that all modifications to the standard model that lead to massive neutrinos change it qualitatively. For one more detailed discussion of this point see, for example,  \cite{DeGouvea:2005gd}.

It is natural to ask what augmented, ``new'' standard model ($\nu$SM) leads to non-zero neutrino masses. The answer is that we are not sure. There are many different ways to modify the standard model in order to accommodate neutrino masses. While most differ greatly from one another, all succeed -- by design -- in explaining small neutrino masses and are all allowed by the current particle physics experimental data. The most appropriate question, therefore, is not what are the candidate $\nu$SM's, but how can we identify the ``correct'' $\nu$SM? The answer lies in next-generation experiments, as will be briefly emphasized in the next section. 

Modifications to the standard model that allow neutrinos to acquire neutrino masses include augmenting the Higgs sector with, say, $SU(2)_L$ Higgs triplets, augmenting the fermion sector with either $SU(2)_L$ singlets (right-handed neutrinos) or triplets, or adding several new fields and interactions that explicitly violate $U(1)_{B-L}$. In order to explain why neutrino masses are so small, other additions are often employed including large extra dimensions, new spontaneously broken gauge symmetries, etc. For detailed examples readers are referred to, for example, \cite{theory_review}. The key point is that different models lead to different phenomenology in the neutrino sector and elsewhere.

\subsection{Understanding Fermion Mixing}

Before concluding, it is worthwhile to mention another perceived neutrino puzzle: the pattern of neutrino mixing. The absolute value of the entries of the CKM quark mixing matrix are, qualitatively, given by
\begin{equation}
|V_{\rm CKM}|\sim \left(\begin{array}{ccc} 1 & 0.2 & 0.001\\ 0.2 & 1 & 0.01 \\ 0.001 & 0.01 & 1\end{array} \right),
\end{equation}
while those of the entries of the PMNS matrix are given by
\begin{equation}
|U_{\rm PMNS}|\sim \left(\begin{array}{ccc} 0.8 & 0.5 & <0.2\\ 0.4 & 0.6 & 0.7 \\ 0.4 & 0.6 & 0.7\end{array} \right).
\end{equation}
It is clear that the two matrices ``look'' very different. While the CKM matrix is almost proportional to the identity matrix plus hierarchically ordered off-diagonal elements, the PMNS matrix is far from diagonal and, with the possible exception of the $U_{e3}$ element, all elements are order one. 

Before lepton mixing was established, naive theoretical expectations were that, if there was indeed lepton mixing, the lepton mixing matrix should ``look like'' the quark mixing matrix. This ``prediction,'' loosely driven by grand unified theories, turned out to be quite wrong. Significant research efforts are concentrated on understanding what, if any, is the relationship between the quark and lepton mixing matrices and what, if any, is the ``organizing principle'' responsible for the observed pattern of neutrino masses and lepton mixing. There are several different theoretical ideas in the market  (for summaries, overviews and references see, for example, \cite{theory_review,Albright:2006cw}). Typical results, which are very relevant for next-generation experiments, include predictions for the currently unknown neutrino mass and mixing parameters ($\sin^2\theta_{13}$, $\cos2\theta_{23}$, the mass hierarchy, etc) and the establishment of ``sum rules'' involving different parameters. 

\section{Where Are We Going? + Conclusions}

Progress in the field of neutrino physics will rely heavily on the availability of new experimental data. These include 
\begin{itemize}
\item the precise determination of ``all'' neutrino oscillation parameters. This is the goal of the next (and next-next) generation of neutrino oscillation experiments \cite{osc_exp};
\item searches for lepton number violation. This is the realm of searches for neutrinoless double-beta decay \cite{absolute_mass};
\item searches for flavor violation in the lepton sector ($\mu\to e\gamma$ \cite{MEG},  $\mu\to e$-conversion in nuclei \cite{mec}, rare tau decays \cite{taus}, etc);
\item precision measurements of charged and neutral lepton electromagnetic properties, like the $g-2$ of the muon, searches for the electron electric dipole moment or searches for neutrino electromagnetic moments \cite{mag_mom}; 
\item searches for the physics responsible for electroweak symmetry breaking and new degrees of freedom at the TeV scale. This is the main purpose of the LHC experiments;
\item Precision measurements of high \cite{Adams:2008cm} and low-energy \cite{sns} neutrino scattering;
\item etc.
\end{itemize} 

In summary, physics beyond the standard model has, after decades of searches, finally manifested itself in one way -- neutrino masses are nonzero! We understand the long-baseline neutrino data very well in terms of neutrino oscillations and have devised a very successful parameterization of the extended neutrino sector. This parameterization allows us to identify what we know we don't know about neutrinos which, in turn, helps define a rich experimental program in neutrino physics for the coming decade or two. 

Theoretically, what we learned can be summarized in two phrases: neutrino masses are very, very small and lepton mixing is very different from quark mixing. We don't have an explanation for either of them but we feel that both of these facts are important clues towards a more satisfying understanding of fundamental physics. 

In order to further our understanding of neutrino masses, lepton mixing and how these $\nu$SM parameters fit into our understanding of fundamental physics, more experimental information is necessary. We are all looking forward to new experimental results which we expect will become available very soon (including LHC results and results from MEG \cite{MEG}).

\begin{acknowledgments}
I am happy to thank the organizers for the invitation to talk about neutrinos to a diverse ``heavy-flavor'' audience. This work is sponsored in part by the US Department of Energy Contract DE-FG02-91ER40684.
\end{acknowledgments}

\bigskip 

\end{document}